# Extracting net current from an upstream neutral mode in the fractional quantum Hall regime

I. Gurman[§], R. Sabo[§], M. Heiblum*, V. Umansky, D. Mahalu

*Braun Center for Submicron Research, Dept. of Condensed Matter physics,*

*Weizmann Institute of Science, Rehovot 76100, Israel*

[§] *Both authors equally contributed to this work*

*\* moty.heiblum@weizmann.ac.il*

# Extracting net current from an upstream neutral mode in the fractional quantum Hall regime

**Upstream neutral modes, counter propagating to charge modes and carrying energy without net charge, had been predicted to exist in some of the fractional quantum Hall states and were recently observed via noise measurements. Understanding such modes will assist in identifying the wavefunction of these states, as well as shedding light on the role of Coulomb interactions within edge modes. In this work, performed mainly in the $\nu = {}^2\!/_3$ state, we placed a quantum dot a few micrometers upstream of an ohmic contact, which served as a 'neutral modes source'. We showed the neutral modes heat the input of the dot, causing a net thermo-electric current to flow through it. Heating of the electrons led to a decay of the neutral mode, manifested in the vanishing of the thermo-electric current at $T > 110 \text{mK}$. This setup provides a straightforward method to investigate upstream neutral modes without turning to the more cumbersome noise measurements.**

## Introduction

Thirty years after the discovery of the celebrated fractional quantum Hall effect (FQHE) [1] there are still open questions regarding electronic characteristics in this regime; one of them is the formation and nature of upstream chiral neutral modes. These modes, which carry energy without net charge [2-10], are at the focus of recent experimental and theoretical studies, and encouraged by the advent of the search for Majorana quasiparticles [11,12]. Certain fractional states, dubbed as hole-conjugate states [13,14], with the most prominent one $\nu = 2/3$, support reconstructed chiral edge modes, which due to interaction and disorder, give birth to counter propagating neutral modes [ 2-4]. Among two possible models of the upstream neutral modes: (*i*) coherent dipole-like excitations, which carry energy; and (*ii*) classical-like excitations, namely, classical heat waves [ 15]; recent results favor the latter one.

Being chargeless, the observation of neutral modes presented a challenge. Bid *et al*. [ 5] allowed upstream neutral modes, emanating from an ohmic contact, to scatter off a partially pinched quantum point contact (QPC). This has resulted with excess noise but, as expected, with zero net current. In a recent work by Gross *et al.* [ 7] a few new features of these modes were reported: first, that a standard (non-ideal) ohmic contact can detect neutral modes very much like a QPC; second, injecting charge onto a QPC also excites upstream neutral modes; and third, a simple model of classical heating was found to explain much of the observed data. Moreover, quantum dots (QDs) were used as thermometers in order to detect temperature increase [ 8,9]. The results, shown in these precursory works, support energy propagation via upstream chiral neutral modes, and exclude transport of lattice phonons.

In this work we follow a recent proposal, which suggested employing a QD to convert the neutral mode's energy into a thermo-electric current [ 10]. We impinged an upstream neutral mode, emanating from the back side of a biased ohmic contact, on the input QPC of a Coulomb blockaded (CB) metallic QD [ 16]. Conductance peaks exhibited substantial broadening when increasing the bias on the ohmic contact. Furthermore, a net thermo-electric current was generated, due to the temperature gradient across the QD. The results agreed with a toy model of a metallic QD having two leads at different temperatures, and the electrons obeying Fermi-Dirac distribution (see more in Supplementary Discussion) [7]. Most of the measurements were carried at $\nu = 2/3$ fractional state, though other filling fractions were also tested.

## Results

### Measurement Setup

The sample was fabricated in a GaAs-AlGaAs heterostructure, embedding a two-dimensional electron gas (2DEG) 130nm below the surface. Schematics of the QD and the measurement setup are depicted in Fig. 1a; accompanied by an SEM micrograph in Fig. 1b. Due to the relatively short decay length of the neutral modes [5], the dot, with internal size $\sim 2 \times 2 \mu m^2$, was placed 10μm upstream of contact N, which served as source for the neutral modes. The current traversing through the dot was darined at contact D, and its RMS value was measured using a spectrum analyzer, thus loosing phase (polarity) information. The sample was cooled to 30mK in a dilution refrigerator.

### Characterizing of the QD

The QD was tuned to the Coulomb blockade regime, having a maximal conductance $G_{\max} \approx \frac{2}{3}\frac{e^2}{h} \times 0.35$ on resonance. The charging energy was deduced via the ubiquitous non-linear differential conductance (the 'Coulomb diamond' in Fig. 1c) [17], $U_c \approx$ 70μeV, and the plunger's levering factor $\alpha_P = \frac{\Delta u}{e \Delta V_P} = 0.005$, where $u$ is the dot's potential. Being relatively large, the QD was metallic, namely, $\delta \ll T \ll U_c$, with $\delta$ the level spacing and $T$ the electron temperature. Therefore, the dot's excitation spectrum was practically continuous (see Fig. 1c), and the width of each conductance peak was expected to be temperature limited [16,18].

**Heating of the QD's Input**

Heating of the QD was monitored by observing the broadening of the conductance peaks in the presence of an upstream neutral mode (red arrow in Fig. 1a). As $V_N$ increased, the conductance peaks broadened (color scale in Fig. 2a) - while their height hardly changed (see Supplementary Fig. S1 & S2). Two cuts, one at $V_N = 0$ and another at $V_N = 200\mu V$ (dashed lines in Fig. 2a), are shown in Fig. 2b. Employing a two-temperature toy model for the QD, with an input temperature $T_{in}$ and an output temperature is 30mK, led to $T_{in} = 125$mK for $V_N = 200\mu V$. A more complete description of $T_{in}(V_N)$ is given in Fig. 2c; it is rather symmetric with respect to $V_N = 0$, with a change in slope around $V_N = 150\mu V$. Such dependence was also observed in previous works [5-8].

The above measurements cannot determine whether the dot was heated homogenously or if a temperature gradient was created. In the most naïve approach, current flows in the dot due to the difference between the non-equilibrium energy distributions in both leads [19]. Therefore, when an external bias is applied, electrons will flow from high to low potential with resultant Coulomb blockage peaks (Fig. 3a). However, the thermo-electric current $I_{te}$, driven solely by a temperature gradient across the dot, will strongly depend on the position of the dot's relevant energy level ($E_D$) with respect to the Fermi energy ($\varepsilon_F$) [20,21]. When $E_D < \varepsilon_F$ electrons will flow from the cold to the hot lead, whereas when $E_D > \varepsilon_F$ they will run in the opposite direction (see Fig. 3b). In Fig. 3c, the blue curve is the measured linear differential conductance, while the red curve is the RMS value of thermo-electric current, when $V_N$ is AC modulated (amplitude of 10µV). More information about the change of sign in $I_{te}$ is presented in Supplementary Figure S3 and Supplementary Discussion. This result proves a thermal gradient was formed across the dot due to an upstream neutral mode.

**Thermo-Electric Current's Dependence on Neutral Energy**

Establishing that the heating of the QD is insensitive to the sign of $V_N$ (Fig. 2) leads to two possible scenarios: - either $I_{te} \propto V_N^2$ or $I_{te} \propto |V_N|$. This was tested by applying $V_N = V_{N,DC} + V_{N,AC}\cos(\omega t)$, scanning $V_{N,DC}$ while keeping $V_{N,AC}$ constant, and measuring the 1st and 2nd harmonics in the current $I_{te}$. If $I_{te} \propto V_N^2 = V_{N,DC}^2 + \frac{1}{2}V_{N,AC}^2 + 2V_{N,DC}V_{N,AC}\cos(\omega t) + \frac{V_{N,AC}^2}{2}\cos(2\omega t)$, then one would expect the 1st harmonic to rise while the 2nd one stays constant with $V_{N,DC}$. Yet, if $I_{te} \propto |V_N|$, the 1st harmonic should increase while the 2nd decrease, to eventually disappear once $V_{N,DC} > V_{N,AC}$. Consequently, the 2nd harmonic is the best criterion to distinguish between the two options. Fig. 4a shows the results of such a measurement with $f = \frac{\omega}{2\pi} = 995(497.5)\text{KHz}$ when measuring the 1st (2nd) harmonic, $V_{N,AC} = 100\mu V$ and $V_{N,DC}$ increasing in small increments. The decrease of the 2nd harmonic (green curve), until its diminishing at $V_{N,DC} = 100\mu V = V_{N,AC}$, clearly suggests that $I_{te}$ is proportional to $|V_N|$ and not to $V_N^2$; as noted in Ref. 15. In addition, the fact that $I_{te}$ followed the ~1MHz excitation on contact N shows that the neutral mode's response time is 1µS at most.

A similar measurement was performed by adding a small AC signal to the plunger gate and varying the DC bias on N. The transconductance $\frac{\partial I_{te}}{\partial V_P}$ was measured at resonance, (where its value is maximal) as function of $V_N$ (this is the actual slope of $I_{te}$ at resonance in Fig. 3c (red curve)). The slope in Fig. 4b is a direct representation of $I_{te}$ as function of $V_N$. The slope moderated near $V_N \sim 150\mu V$; attributed to a high enough temperature increase, which may allow two electrons to traverse through the QD - each contributes to an opposite polarity current - leading to the apparent saturation.

## Discussion

After establishing a clear indication of heating by the neutral mode, we investigated its dependence on the base temperature. Kane *et al.* [3,4] predicted that the energy carried by the upstream neutral modes decays exponentially with distance, with a characteristic decay length proportional to $T^{-2}$, namely, as $exp\left\{-\left(\frac{T}{T_0}\right)^2\right\}$, with $T_0$ a characteristic temperature. In the current setup, raising the lattice temperature (by heating the mixing chamber of the dilution refrigerator, $T_{MC}$) should increase the temperature of the output QPC while lowering the temperature rise at the input QPC. Consequently, the temperature gradient across the QD reduces rapidly, and with it the thermo-electric current should ebb.

Fig. 5a shows $T_{in}$ (deduced in a similar fashion to that described before, see Fig. 2c) as function of $V_N$ at different values of $T_{MC}$. It is clear that as $T_{MC}$ increases the temperature rise of the hot input (above $T_{MC}$) becomes smaller; without further increase for $T_{MC} >$ 110mK. Via standard noise measurements we verified that at this base temperature the neutral signal decreased substantially [ 5]. These results clearly point out a strong decay of the neutral mode with temperature.

Furthermore, we measured the maximal (as function of $V_P$) thermo-electric current $I_{te}$ for different values of $T_{MC}$, while keeping a constant excitation of 300μV on $V_N$. The data (open circles in Fig. 5b) was compared with a simulation of $I_{te}$ based on a hot input temperature $T_{in} = T_{MC} + \alpha e^{-\left(\frac{T_{MC}}{T_0}\right)^2}$ and a cold output (at $T_{MC}$). We found that $\alpha =$ 372mK and $T_0 = $ 66mK lead to an excellent agreement with the data (solid red line in Fig. 5b); thus supporting the predicted temperature dependence [ 3,4].

Similar measurements were performed at $\nu = 3/5$, being also a hole-conjugate state, with essentially the same qualitative behavior [ 5]. However, no upstream heating had been observed at $\nu = 1$ and $\nu = 2$ [ 22] (as shown in Supplementary Figure S4); confirming that the observed heating is due to the presence of upstream neutral modes at certain fractional states.

In this work, we demonstrated a new means – exploiting a QD - to transform energy carried by upstream neutral modes into net current; a method that is likely to be simpler than measuring noise. Employing a smaller QD, with a quantized excitation spectrum ($\delta > T$), is expected to provide an experimental gateway to further understanding other properties of the neutral modes [10]. Furthermore, the measurement techniques reported in this work can be used to explore the nature of the neutral modes at the $\nu = 5/2$ state [5, 6], which is suspected to carry with it non-Abelian braiding statistics [23].

## Methods

### Sample Fabrication

The sample was fabricated in a GaAs-AlGaAs heterostructure, embedding a two-dimensional electron gas (2DEG), with areal density $8.8 \times 10^{10} \text{cm}^{-2}$ and 4.2K 'dark' mobility $5.8 \times 10^6 \text{cm}^2\text{V}^{-1}\text{s}^{-1}$, 130nm below the surface. Schematics of the QD and the measurement setup are depicted in Fig. 1a; accompanied by an SEM micrograph in Fig. 1b. The dot, with internal size $\sim 2 \times 2 \mu\text{m}^2$, was defined by two split metallic gates (TiAu), serving as QPCs with openings 650nm wide. A plunger gate, 300nm wide, controlled the occupation of the dot. Due to the relatively short decay length of the neutral modes [5], contact N, serving as source for the neutral modes, was placed 10µm downstream the QD. Contacts C, D, G1 & G2 were placed tens of micrometers away from the dot. The sample was cooled to 30mK in a dilution refrigerator.

### Measurement Technique

Biasing contact C raised the chemical potential of the input QPC of the QD, while biasing contact N increased the QPC temperature (see later). In both cases, a net electrical current $I$ traversed the dot. The output drain (D) voltage $V_\text{D} = IR_\text{H}$, with $R_\text{H}$ the Hall resistance, was filtered using an LC resonant circuit ($f_0 = 995$ KHz) and amplified by homemade voltage preamplifier (cooled to 1K) followed by a room temperature amplifier (NF SA-220F5). A commercial spectrum analyzer displayed the RMS signal; hence, loosing phase (polarity) information.

# References

1. Tsui, D. C., Stormer, H. L. & Gossard, A. C., Two-Dimensional Magnetotransport in the Extreme Quantum Limit. *Phys. Rev. Lett.* **48**, 1559-1562 (1982).

2. MacDonald, A. H., Edge states in the fractional quantum Hall effect regime. *Phys. Rev. Lett.* **64**, 220-223 (1990).

3. Kane, C. L. & Fisher, M. P. A., Impurity scattering and transport of fractional quantum Hall edge states. *Phys. Rev. B* **51**, 13449-13466 (1995).

4. Kane, C. L., Fisher, M. P. A. & Polchinski, J., Randomness at the Edge: Theory of Quantum Hall Transport at Filling ν=2/3. *Phys. Rev. Lett.* **72**, 4129-4132 (1994).

5. Bid, A. *et al.*, Observation of Neutral Modes in the Fractional Quantum Hall Regime. *Nature* **466**, 585-590 (2010).

6. Dolev, M. *et al.*, Characterizing Neutral Modes of Fractional States in the Second Landau Level. *Phys. Rev. Lett.* **107**, 036805 (2011).

7. Gross, Y., Dolev, M., Heiblum, M., Umansky, V. & Mahalu, D., Upstream neutral modes in the fractional quantum Hall effect regime: heat waves or coherent dipoles? *Phys. Rev. Lett.* **108**, 226801 (2012).

8. Venkatachalam, V., Harty, S., Pfeiffer, L., West, K. & Yacoby, A., Local Thermometry of Neutral Modes on the Quantum Hall Edge. *Nature Physics* **8**, 676-681 (2012).

9. Altimiras, C. *et al.*, Chargeless heat transport in the fractional quantum Hall regime. *Phys. Rev. Lett.* **109**, 026803 (2012).

10. Viola, G., Das, S., Grosfeld, E. & Stern, A., Thermoelectric probe for neutral edge modes in the fractional quantum Hall regime. *Phys. Rev. Lett.* **109**, 146801 (2012).

11. Lee, S. S., Ryu, S., Nayak, C. & Fisher, M. P. A., Particle-Hole Symmetry and the 5/2 Quantum Hall State. *Phys. Rev. Lett.* **99**, 236807 (2007).

12. Yu, Y., Dynamics of edge Majorana fermions in fractional quantum Hall effects. *J. Phys.: Condens. Matter* **19**, 466213 (2007).

13. MacDonald, A. H. & Rezayi, E. H., Fractional quantum Hall effect in a two-dimensional electron-hole fluid. *Phys. Rev. B* **42**, 3224-3227 (1990).

14. Girvin, S. M., Particle-hole symmetry in the anomalous quantum Hall effect. *Phys. Rev. B* **29**, 6012-6014 (1984).

15. Takei, S. & Rosenow, B., Neutral mode heat transport and fractional quantum Hall shot noise. *Phys. Rev. B* **84**, 235316 (2011).

16. van Houten, H., Beenakker, C. W. J. & Staring, A. A. M., in *Single Charge Tunneling*, edited by Grabert, H. & Devoret, M. H. (Plenum Press, New York, 1992), pp. 167-216.

17. Reimann, S. M. & Manninen, M., Electronic structure of quantum dots. *Rev. Mod. Phys.* **74**, 1283-1342 (2002).

18. Beenakker, C. W. J., Theory of Coulomb-blockade oscillations in the conductance of a quantum dot. *Phys. Rev. B* **44**, 1646-1656 (1991).

19. Altimiras, C. *et al.*, Non-equilibrium edge-channel spectroscopy in the integer quantum Hall regime. *Nature Physics* **6**, 34-39 (2010).

20. Beenakker, C. W. J. & Staring, A. A. M., Theory of the thermopower of a quantum dot. *Phys. Rev. B* **46**, 9667-9676 (1992).

21. Staring, A. A. M. *et al.*, Coulomb-Blockade Oscillations in the Thermopower of a Qauntum Dot. *Europhys. Lett.* **22** (1), 57-62 (1993).

22. Granger, G., Eisenstein, J. P. & Reno, J. L., Observation of Chiral Heat Transport in the Quantum Hall Regime. *Phys. Rev. Lett.* **102**, 086803 (2009).

23. Read, N. & Green, D., Paired states of fermions in two dimensions with breaking of parity and time-reversal symmetries and the fractional quantum Hall effect. *Phys. Rev. B* **61**, 10267-10297 (2000).

**Acknowledgments**

The authors will like to acknowledge O. Zilberberg, Y. Gross, A. Stern and G. Viola for fruitful discussions and remarks on the manuscript. We acknowledge the partial support of the European Research Council under the European Community's Seventh Framework Program (FP7/2007-2013) / ERC Grant agreement # 227716, the Israeli Science Foundation (ISF), the Minerva foundation, the German Israeli Foundation (GIF), the German Israeli Project Cooperation (DIP), and the US-Israel Bi-National Science Foundation (BSF). I.G. is grateful to the Azrieli Foundation for the award of an Azrieli Fellowship.

## Author Contributions

I.G, R.S and M.H designed and preformed the experiment and wrote the paper; V.U grew the 2DEG and D.M preformed the electron beam lithography.

## Competing Financial Interests

The authors declare no competing financial interests.

# Figures

Figure 1:

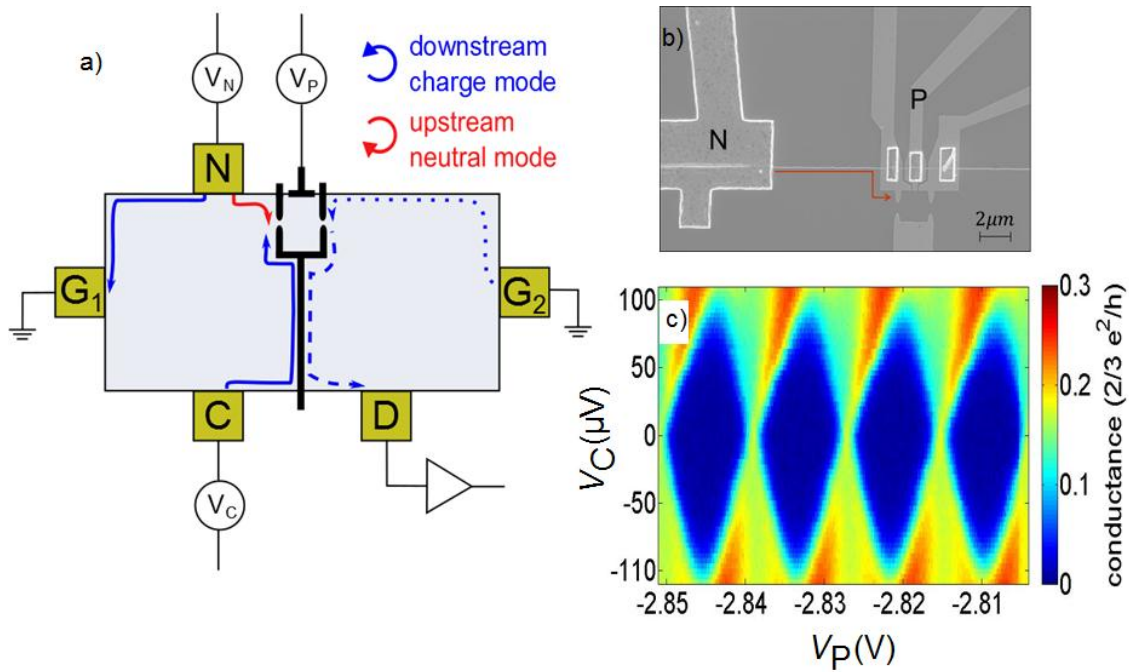

Quantum dot scheme:

a) A sketch of the device. Either a charge current injected from contact C (blue) or an upstream neutral current injected from contact N (red) are impinged on the input of the QD. The current at the output of the QD (dashed blue) is measured at contact D through an LC ($f_0 = 995$KHz) circuit and a home-made cold amplifier. A constant flow of charged mode fixed at ground potential (contact G2) is impinged on the other side of the dot (blue dotted line). The charge mode emitted from contact N flows directly into the ground (G1, blue) and influences neither the QD nor the experiment.

b) SEM image showing the QD and the ohmic contact used to inject the neutral mode. The upstream neutral current travels a distance of 10µm along the edge before impinging at the QD (red arrow). All the other ohmic contacts (shown only in (a)) are tens of microns away from the dot. The QD itself is made out of two QPCs (opening of 650nm) and a plunger gate (300nm wide). The lithographic area of the QD is $\sim 2 \times 2 \mu m^2$.

c) Coulomb diamond characterization of the QD in $\nu = 2/3$, obtained by measuring the QD's conductance, $\frac{\partial I}{\partial V_C}$, while scanning $V_P$ and changing $V_C$ in small DC increments. The scale bar is in units of quantum conductance at $\nu = 2/3$ - $G(2e^2/3h)$.

Figure 2:

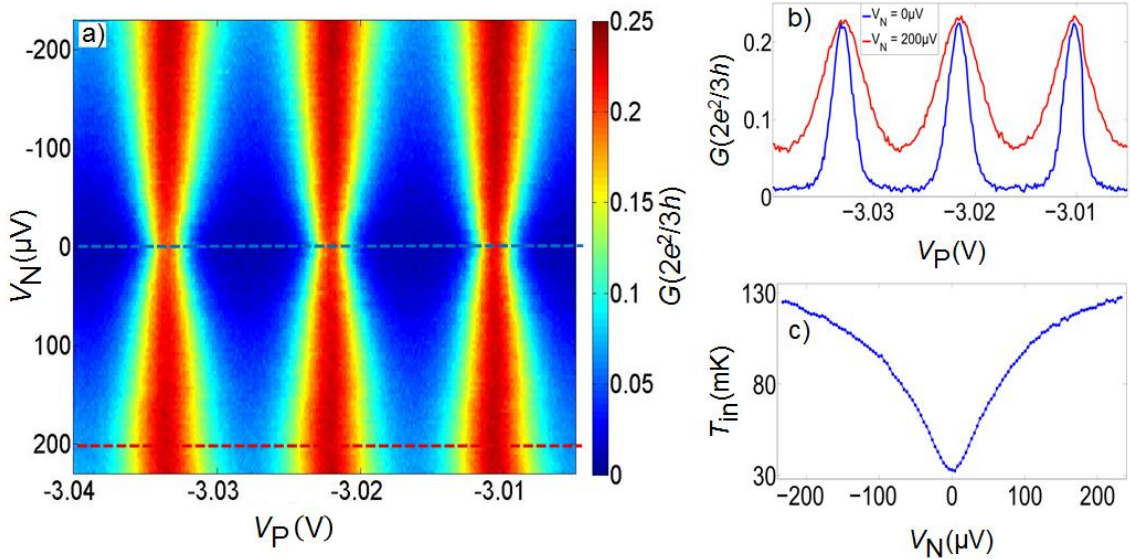

Broadening of CB peaks due to neutral mode:

a) 2D plot showing the broadening of three consecutive Coulomb blockade peaks and serving as a proof for the heating caused by the neutral mode.

b) Two cuts of the 2D plot showing Coulomb blockade peaks at $V_N = 0$ (blue) and $V_N = 200\mu V$ (red). The dashed lines in (a) show the lines along which the cuts were made.

c) The hot lead's temperature ($T_{in}$) as function of $V_N$. The values were deduced from the width of the CB peaks using a toy model of a metallic QD with different temperatures at its leads.

Figure 3:

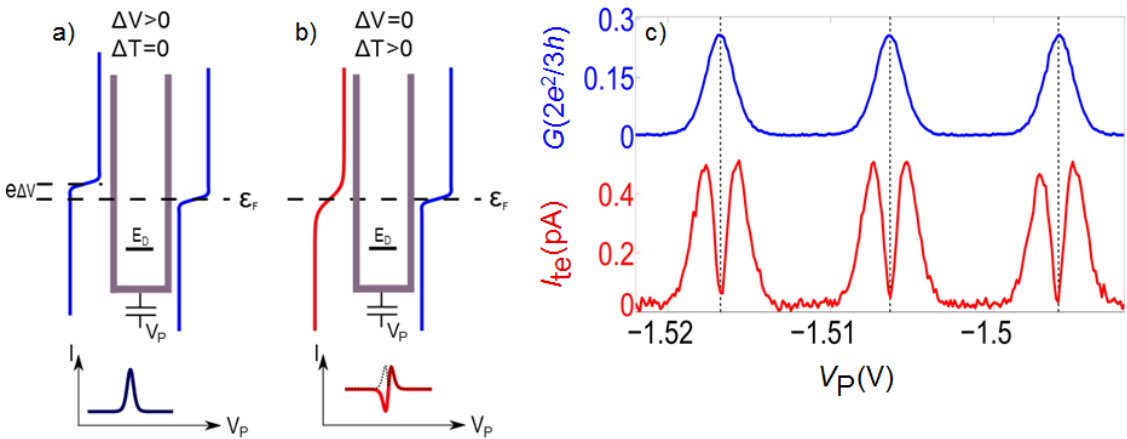

Current flow due to either voltage or temperature gradient:

An illustration of a quantum dot under electrical or thermal gradient, showing how the current behaves once the QD is near resonance. The energy levels are capacitively coupled to the plunger gate ($V_P$), and for simplicity only one energy level ($E_D$, marked by a full line) is shown. (a) Applying a finite voltage on the left lead will result in current from left to right once the energy level is in the vicinity of the leads' Fermi energy ($\varepsilon_F$, marked by dashed line); hence the single peak shown. The blue curve in (c) shows the measured conductance peaks, which fit this picture. (b) Heating up the left lead will result in current from right to left as $E_D < \varepsilon_F$, zero current at $E_D = \varepsilon_F$ and current from left to right when $E_D > \varepsilon_F$. We also show the absolute value of the thermo-electric current (dotted black line) to match our measuring capabilities. The red curve in (c) shows the thermo-electric current due to voltage applied on contact N. The measurement gives a good agreement with the thermal gradient picture.

Figure 4:

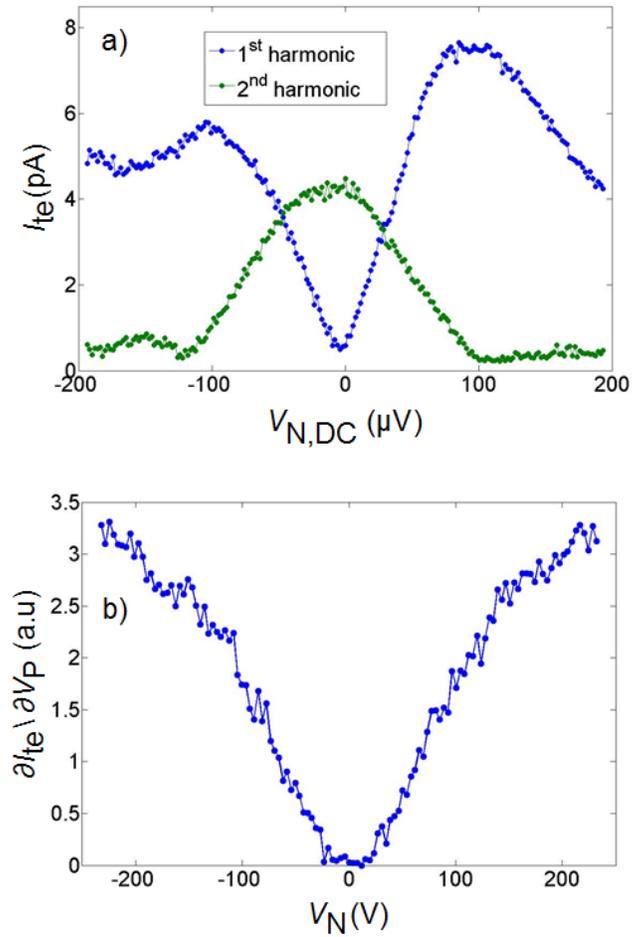

Dependence of the resulting thermo-electric signal on the neutral voltage:

a) Change of the thermo-electric current's first two harmonics when applying $V_N = V_{N,DC} + V_{N,AC}\cos(\omega t)$ ($V_{N,AC} \sim 100\mu V$). As the first harmonic (blue line) increases so does the second harmonic (green line) drops. This outcome proves that the thermo-electric current $I_{te}$ is proportional to $|V_N|$.

b) Amplitude of the thermal transconductance as function of the DC bias on $V_N$ and a constant AC modulation on the plunger gate.

Figure 5:

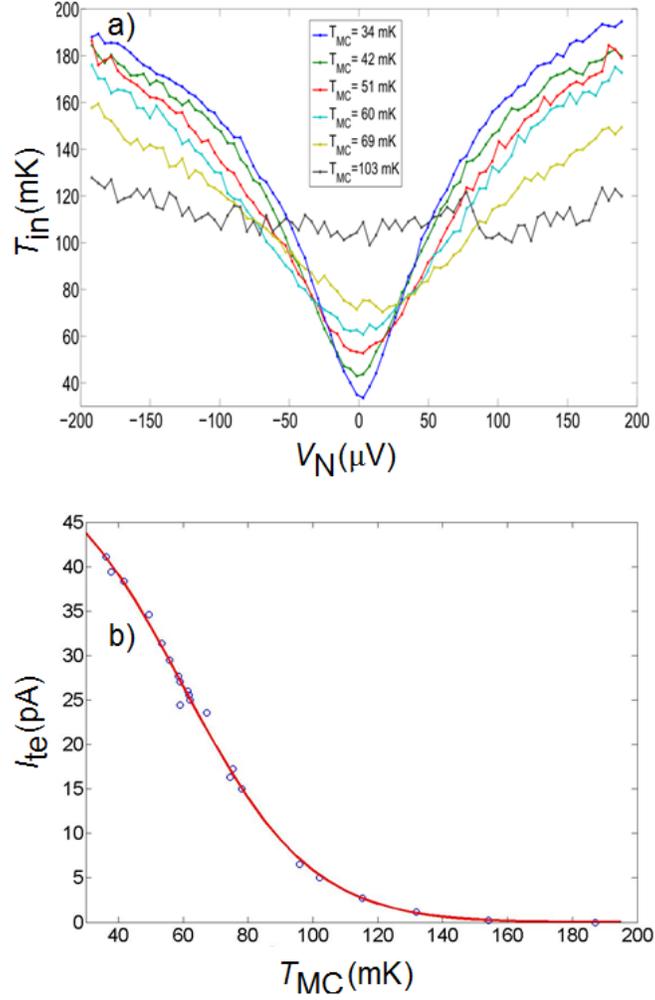

Temperature dependence of the neutral modes' energy:

a) Hot lead's temperature ($T_{in}$) as function of the DC bias on contact N for different base temperatures (34mK − 103mK). As the temperature increases the neutral mode's energy decreases leading to weaker heating of the QD's input.

b) Thermo-electric current dependence on base temperature, while applying a constant AC signal on contact N. The data is fitted to a quantum dot with a temperature gradient, where $\Delta T = \alpha\, exp\left\{-\left(\frac{T_{MC}}{T_0}\right)^2\right\}$. The data (blue circles) is very well described (red line) by our toy model using $\alpha = 372$mK and $T_0 = 66$mK.

# Supplementary Information for Extracting net current from an upstream neutral mode in the fractional quantum Hall regime

# **Supplementary Figures**

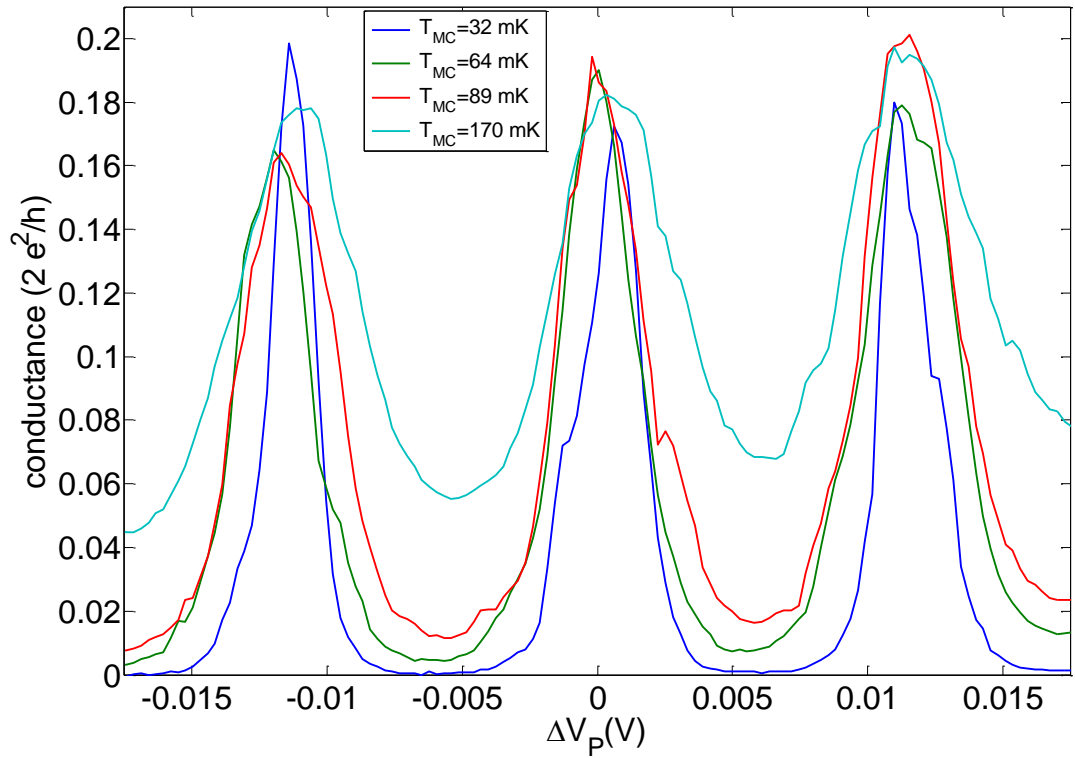

**Supplementary Figure S1 - CB peaks at $\nu = 2$ for different values of $T_{MC}$.** No significant effect of the temperature on the height of the peaks is observed.

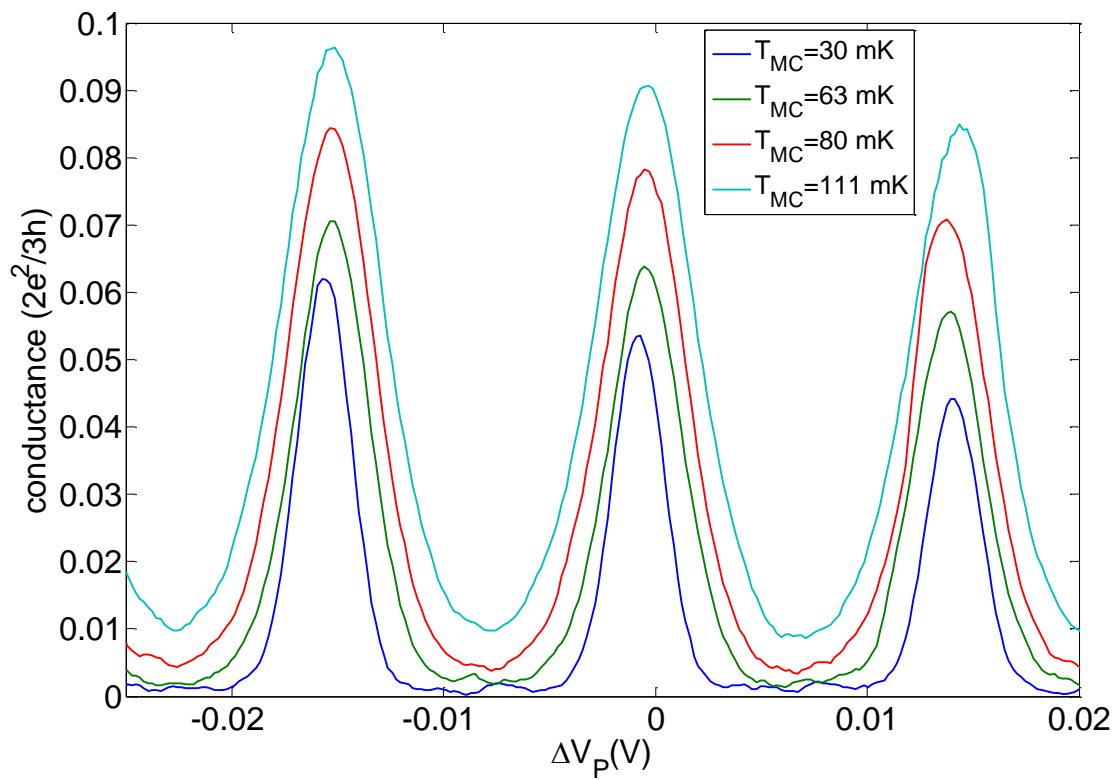

**Supplementary Figure S2 - CB peaks at $\nu = {}^2\!/_3$ for different $T_{MC}$.** A clear increase of the amplitude of the CB peaks when raising the temperature is observed.

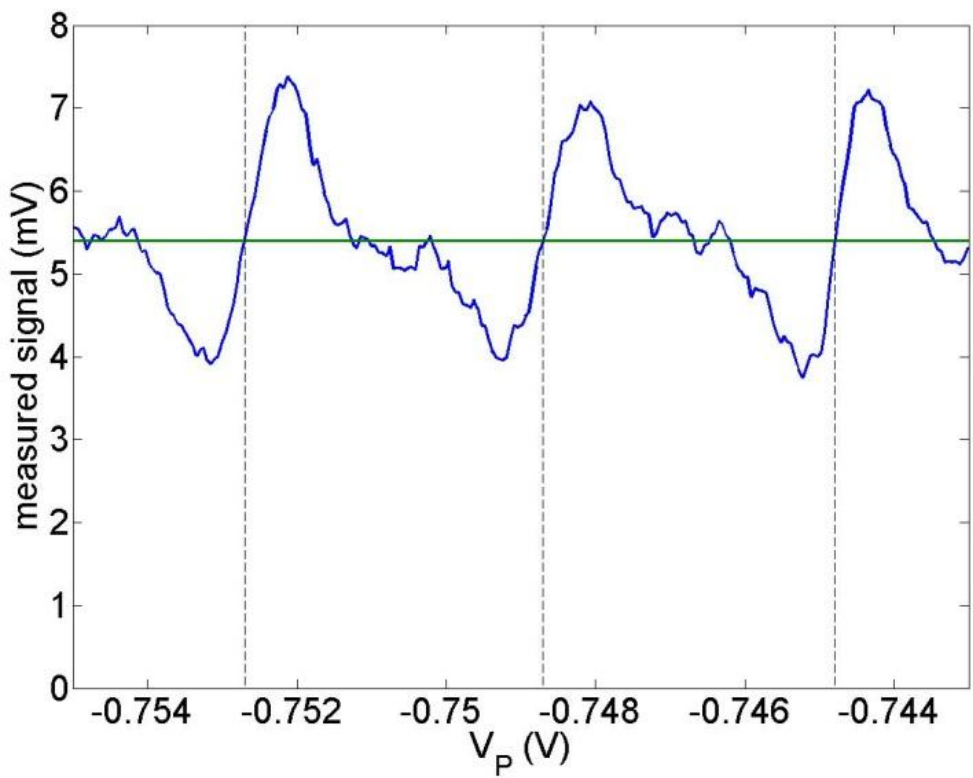

**Supplementary Figure S3 – thermo-electric current with high background shows the current changes polarity.** Thermoelectric current due to AC signal on contact N. The different contributions of the signal around the background (green line), prove that the current changes its sign at exact resonance (dashed line).

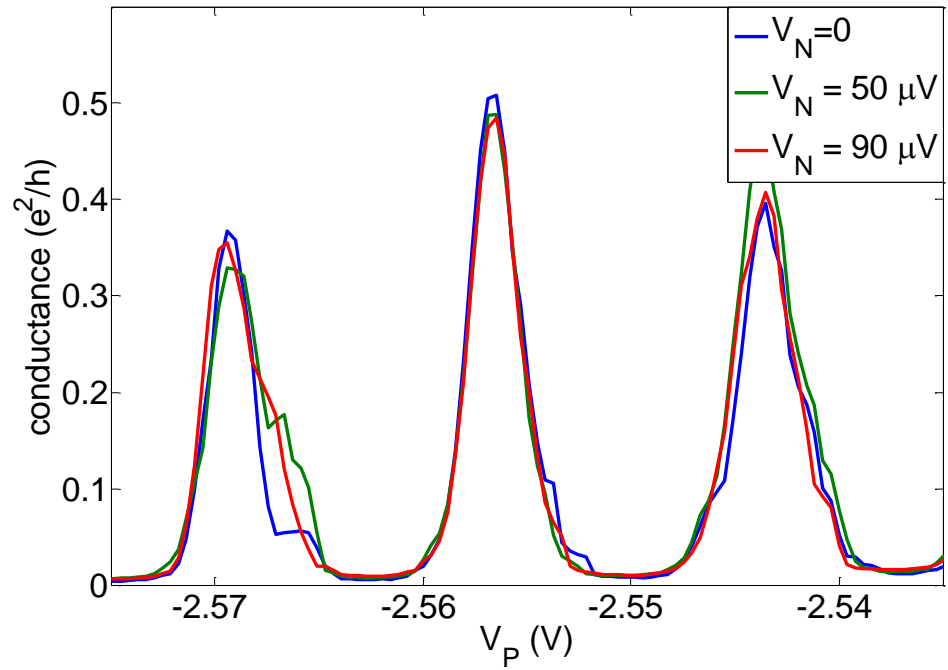

**Supplementary Figure S4 - CB peaks at $\nu = 2$ for different $V_N$.** Comparison with Fig. 2b of the main text suggests that the heating observed in $\nu = 2/3$ depends on the state of 2DEG, thus support the presence of upstream neutral modes.

# SUPPLEMENTARY NOTES

**Rigorous calculation of a metallic quantum dot with temperature and chemical potential gradients**

It is quite known that the electrons which move along the edges of fractional quantum Hall states cannot be described using the regular Fermi-Dirac (FD) distribution. However, due to simplicity considerations we chose to describe our system via the FD distribution and use it in order to conduct all the following calculations.

In this part we are going to fully derive the formula from which we computed the temperature of the hot lead. Here we followed chapter 10.2 from the book "Many-body quantum theory in condensed matter physics: an introduction", by H. Bruss and K. Flensberg, Oxford University Press (2004).

The energy levels of a metallic dot depend solely on Coulomb interactions and the influence of the plunger gate –

$$E(N) = E_\text{C} N^2 - eV_\text{P} N = \frac{e^2 N^2}{2C} - eV_\text{P} N = \frac{1}{2C}(eN - V_\text{P}C)^2 - \frac{V_\text{P}^2 C}{2} \qquad \text{(S1)}$$

where : $E_\text{C}$ – charging energy, $N$ – number of electrons in the dot, $V_\text{P}$ – plunger voltage, $e$ – electron charge, $C$ – capacitance of the dot.

According to Fermi's golden rule, tunneling rate through the left lead of the quantum dot is –

$$\Gamma^L_{\beta\alpha} = 2\pi \sum_{f_\beta, i_\alpha} |\langle f_\beta | H_{TL} | i_\alpha \rangle|^2 W_{i_\alpha} \delta\left(E_{f_\beta} - E_{i_\alpha}\right) \tag{S2}$$

where : $i_\alpha(f_\beta)$ – initial (final) state of the system, $H_{TL}$ – tunneling Hamiltonian for the left lead, $W_{i_\alpha}$ – thermal distribution function, $E_{i_\alpha}\left(E_{f_\beta}\right)$ – initial (final) energy.

Same expression holds for tunneling through the right lead.

To simplify $\Gamma^L_{\beta\alpha}$ we first look at the event of an electron tunneling into the dot, such that –

- $i_\alpha = i_N$
- $f_\beta = f_{N+1} \Rightarrow |f_\beta\rangle = c^\dagger_D c_L |i_\alpha\rangle$
- $H_{TL} = t_{LD} c^\dagger_D c_L$
- $E_{i_\alpha} = \varepsilon_L + E(N)$
- $E_{f_\beta} = \varepsilon_D + E(N+1)$

where: $c^\dagger_i$ ($c_i$) – creation (annihilation) operator, $t_{LD}$ - tunneling energy from lead to dot, $\varepsilon_i$ - energy level of the electrons.

Therefore we get –

$$\Gamma^{\text{L}}_{N+1,N} = 2\pi \sum_{D,L} \sum_{i_N} |t_{\text{LD}}|^2 |\langle i_N | c_{\text{L}}^\dagger c_{\text{D}} c_{\text{D}}^\dagger c_{\text{L}} | i_N \rangle|^2 W_{i_N} \delta[\varepsilon_{\text{D}} - \varepsilon_{\text{L}} + E(N+1) - E(N)]$$
(S3)

In addition we assume –

- the lead and dot states are independent - $W_{i_\alpha} = W_{i_L} W_{i_{ND}}$ and $|i_N\rangle = |i_L\rangle \otimes |i_{ND}\rangle$
- tunneling density of states is constant in the relevant energy range – we can take sum over $|t_{\text{LD}}|^2$ separately and get a $\gamma_L$ factor to the whole expression.
- the energy levels are continuous, such that the sum turns into an integral.
- the electrons in the dot and the leads are described by the Fermi-Dirac distribution $n_F^i(\varepsilon_i) = \left(1 + e^{\varepsilon_i \backslash k_B T_i}\right)^{-1}$, where the lead and dot can have different temperatures and chemical potentials.

Under these assumptions we get –

$$\sum_{i_{ND}} W_{i_{ND}} |\langle i_{ND} | c_D c_D^\dagger | i_{ND} \rangle|^2 = 1 - n_F^D (\varepsilon_D - \mu_D)$$

$$\sum_{i_L} W_{i_L} |\langle i_L | c_L^\dagger c_L | i_L \rangle|^2 = n_F^L (\varepsilon_L - \mu_L)$$
(S4)

where: $\mu_i$ - chemical potential of the dot\lead.

Obviously, all this process can be done for the right lead. Therefore, we finally get the final expression for the tunneling rate of an electron from lead $\alpha$ into a dot with $N$ electrons:

$$\Gamma^\alpha_{N+1,N} = \gamma_\alpha \int_{-\infty}^{\infty} d\varepsilon_D d\varepsilon_\alpha n_F^\alpha(\varepsilon_\alpha - \mu_\alpha)[1 - n_F^D(\varepsilon_D - \mu_D)]\delta[\varepsilon_D - \varepsilon_\alpha + E(N+1) - E(N)]$$

$$= \gamma_\alpha \int_{-\infty}^{\infty} d\varepsilon \, n_F^\alpha[E(N+1) - E(N) + \mu_D - \mu_\alpha + \varepsilon] \times [1 - n_F^D(\varepsilon)]$$

(S5)

In the same manner we can get the rate of electron tunneling from a dot with N electrons to lead $\alpha$ –

$$\Gamma^\alpha_{N-1,N} = \gamma_\alpha \int_{-\infty}^{\infty} d\varepsilon \, n_F^D[E(N-1) - E(N) - \mu_D + \mu_\alpha + \varepsilon] \times [1 - n_F^\alpha(\varepsilon)] \qquad (S6)$$

The next step is to find the probability for the dot to have $N$ electrons under the assumption that the dot is in a state such that it can have only $N$ or $N-1$ electrons in it. We neglect the other possibilities as we assume, for the benefit of this calculation, the temperature to be smaller than the charging energy. Using rate equations we get –

$$\Gamma_{N,N-1} P(N) = \Gamma_{N-1,N} P(N-1)$$

(S7)

$$P(N) + P(N-1) = 1$$

where : $\Gamma = \Gamma^L + \Gamma^R$

Finally we get the total current through the dot to be –

$$I = -e \sum_n \left(\Gamma^L_{n+1,n} - \Gamma^L_{n-1,n}\right) P(n) = e\left[P(N-1)\Gamma^L_{N,N-1} - P(N)\Gamma^L_{N-1,N}\right]$$

$$= e\left[\frac{\Gamma^L_{N,N-1}}{1 + \frac{\Gamma^L_{N,N-1} + \Gamma^R_{N,N-1}}{\Gamma^L_{N-1,N} + \Gamma^R_{N-1,N}}} - \frac{\Gamma^L_{N-1,N}}{1 + \frac{\Gamma^L_{N-1,N} + \Gamma^R_{N-1,N}}{\Gamma^L_{N,N-1} + \Gamma^R_{N,N-1}}}\right] \quad (S8)$$

Supplementary Equation S8 is the most general expression for current via a metallic quantum dot with different temperatures and chemical potentials in the leads and dot. From this expression we can deduce many measurable sizes, such as the zero bias conductance or the thermo-electric current solely due to temperature gradient.

The QD input's temperatures displayed in the paper were extracted via comparison between our measurements to numerical calculations of the current (Supplementary Equation S8). In the simulations we assumed the cold output and the dot itself to be in the same temperature (base temperature), for the tunneling electrons do not go through any relaxation process inside the dot.

When assuming the dot's temperature is an average between the temperatures of the two leads, the input's extracted temperature decreases, yet not in a manner that will change the picture presented in the main text.

**Temperature dependence of the Coulomb blockade peaks at different filling factors**

In the main text we have described the behavior of the Coulomb blockade (CB) peaks when we sent energy carried by the upstream neutral to the quantum dot (QD). We have noted that the energy carried by the neutral mode broadened the CB peaks while keeping their amplitude fixed. In the section we will describe the behavior of the CB peaks when we heat up our system by raising the mixing chamber's temperature ($T_{\text{MC}}$).

Starting with the quantum Hall state $\nu = 2$, we show in Supplementary Figure S1 that as we increase $T_{\text{MC}}$ from 32mK to 170mK the CB peak broadens yet there is no significant effect on the amplitude at resonance. This behavior is the expected from a metallic QD when the leads and dot can be describe by a Fermi liquid theory.

Completely different results were obtained at the $\nu = 2/3$ Hall state. Here, as seen in Supplementary Figure S2, heating up $T_{\text{MC}}$ from 30mK to 111mK clearly increased the value of the conductance at resonance. This behavior can only be explained if we leave the framework of Fermi liquid theory.

Furusaki [24] has shown that in the case of chiral Luttinger liquids (CLL) with $g < 1/2$ the amplitude of the CB peaks should increase with temperature. Although the $\nu = 2/3$ state cannot be described as CLL, its hole-conjugate $\nu = 1/3$ state is a CLL with $g = 1/3$. This outcome points to the conclusion that the edge states of $\nu = 2/3$ cannot be fully depicted using Fermi liquid, but in fact require a more subtle treatment.

**The opposite sign of $I_{te}$**

In this part we will show that the two different peaks in the thermoelectric current are indeed with opposite signs. The thermo-electric current shown in Fig. 3c of the main text is measured by applying an AC signal on contact N at frequency $f = f_0/2 = 497.5 \text{KHz}$ and measuring the resulting $I_{te}$ at a frequency of $f = f_0$. A second approach is to use the same frequency $f_0$ for both the applied at the measured signals. In the latter case, there is a need to apply a DC bias on contact N in order to measure the thermoelectric response (see Fig. 4a in the main text).

In the case of an applied signal at $= f_0/2$ , we are sensitive only to a signal with double the frequency. The only mechanism in our setup that would double the frequency is the fact that the rise in temperature due to neutral mode is sensitive to the absolute value of the applied voltage. Therefore there is hardly any background added to $I_{te}$. The lack of significant background is useful as we are sensitive to small signals yet insensitive to their sign, for the spectrum analyzer measures only RMS.

In the second case, an applied signal at $f = f_0$ induces a significant constant background at the output due to unavoided cross-talk between the wires. The added $I_{te}$ will interfere with the background either constructively or destructively depending on the sign of $I_{te}$: a positive current will increase the measured signal while a negative one will decrease it.

Supplementary Figure S3 shows the signal measured at the spectrum analyzer around three consecutive resonances (their locations are marked with horizontal dashed lines). The opposite sign of $I_{te}$ in each side of the resonance is clear when comparing it to the background level (green line). This result acts as another proof for that $I_{te}$ changes its sign when crossing the resonance peak.

**Observing of null heating at integer filling factors**

As mentioned in the main paper, the experiments done at $\nu = {}^2/_3$ were also conducted in integer filling factors. This was done in order to rule out different heating mechanisms other than upstream neutral modes, for example – phonons.

All the measurements preformed in integer filling factors showed no sign of heating. As an example, we present the results of a measurement done in the $\nu = 2$ state, which is similar to the one shown in Fig. 2b of the main text - where a clear broadening of the CB peaks was observed when DC bias on contact N was applied.

As can be seen in Supplementary Figure S4, no sign of broadening of the CB peaks was observed when the bias on contact N was raised from 0 to 50µV and 90µV. This result, compared to the clear signature exhibited in $\nu = {}^2/_3$, shows that the mechanism behind the heating is related to the specific state of the 2DEG, thus strongly supports the presence of upstream neutral modes over other scenarios. Specifically, lattice related explanations (phonons) are ruled out.

# **Supplementary References**

[24] Furusaki, A. Resonant tunneling through a quantum dot weakly coupled to quantum wires or quantum Hall edge states. *Phys. Rev. B* **57**, 7141–7148 (1998)
35